\documentclass[review]{elsarticle}

\journal{Optics Communications}

\begin{document}

\begin{frontmatter}

\title{Compact and low-cost structured illumination microscopy using an optical fiber coupler}


\author[mymainaddress,mysecondaryaddress]{Shiming Hu}
\author[mymainaddress,mysecondaryaddress]{Lei Liu}
\author[mymainaddress,mysecondaryaddress]{Yizheng Huang}
\author[mymainaddress]{Wenwen Liu}
\author[mymainaddress]{Qingquan Wei}
\author[mymainaddress,mysecondaryaddress]{Manqing Tan}

\author[mymainaddress,mysecondaryaddress]{Yude Yu\corref{mycorrespondingauthor}}
\cortext[mycorrespondingauthor]{Corresponding author}
\ead{yudeyu@semi.ac.cn}

\address[mymainaddress]{State Key Laboratory of Integrated Optoelectronics, Institute of Semiconductors, Chinese Academy of Sciences, Beijing 100083, China}
\address[mysecondaryaddress]{College of Materials Science and Opto-Electronic Technology, University of Chinese Academy of Sciences, Beijing 100049, China}

\begin{abstract}
In this paper, a compact and low-cost structured illumination microscope (SIM) based on a $2 \times 2$ fiber coupler is presented. Fringe illumination is achieved by placing two output fiber tips at a conjugate Fourier plane of the sample plane as the point sources. Raw structured illumination (SI) images in different pattern orientations are captured when rotating the fiber mount. Following this, high resolution images are reconstructed from no-phase-shift raw SI images by using a joint Richardson-Lucy (jRL) deconvolution algorithm. Compared with other SIM setups, our method provides a much shorter illumination path, high power efficiency, and low cost.
\end{abstract}


\end{frontmatter}


\section{Introduction}

Fluorescence microscopy plays a vital role in modern biological research owing to its non-invasive visualization of biological samples. However, due to its optical diffraction limit, the lateral resolution of a conventional fluorescence microscope is restricted at $\lambda / (2NA)$, where $\lambda$ is the wavelength of the light source and $NA$ is the numerical aperture of the objective lens \cite{Abbe1873}. In order to obtain fine structures at sub-diffraction sizes, several super-resolution (SR) imaging techniques have been developed to bypass resolution barriers \cite{Gustafsson2000,HELL1994,Rust2006,Betzig2006}. Among these SR techniques, structured illumination microscopy (SIM) stands out owing to its simple optical configuration and wide-field full-frame images at high acquirement speeds \cite{Dan2014}.

In a typical SIM setup, sinusoidal patterns are used to excite the sample fluorescence, which can shift the high-frequency component of the sample structure into the passband of the optical imaging system; resolution-enhanced images could be recovered from multiple SI image acquisitions with at least three phase shifts. Isotropic SR results can be obtained by repeating this process at other orientations. Thus, several devices have been proposed to generate those structured illuminations: gratings \cite{Wang2011}, diffractive optical elements (DOEs) \cite{Rodriguez2008}, spatial light modulators (SLMs) \cite{Chang2009,Lu-Walther2015,Song2016}, digital micro-mirror device (DMD) \cite{Dan2013}, silicon-on-insulator (SOI) chips \cite{Liu2016}, and plasmonic nanostructured substrates \cite{Wei2014}. Among these devices, the SLM is the most widely used for building a high-performance SIM system that can provide a flexible and accurate way to achieve phase, frequency, and orientation modulation of illumination fringes. However, its prohibitive cost and low power efficiency owing to the diffraction of grating structures limited the penetration of SIM technique.

In recent years, it has been proposed that SR images can be recovered from three raw SI images by using a maximum likelihood approach \cite{Str?hl2017} with a joint Richardson-Lucy (RL) deconvolution algorithm \cite{Ingaramo2014,Strohl2015}, meaning the phase modulation of the illumination pattern is no longer required for SIM image processing.

In this paper, we propose a simple and low-cost method to achieve structured illumination super-resolution imaging using a $2\times2$ fiber coupler, wherein two output fiber tips were mounted in parallel on a nested rotation mount modified with a 3D-printed component. By placing this fiber mount at the conjugate Fourier plane of the illumination path, interference illumination can be obtained at the sample plane. High resolution images were recovered from no-phase-shift raw SI images (one SI image per orientation) by using a joint Richardson-Lucy deconvolution algorithm. Our method offers an inexpensive way to build a compact SIM system with high power efficiency.

\section{Method}

In mathematics, an illumination field of two-beam interference can be expressed as:
\begin{equation}
E(r) = cos(\omega_0 r + \phi), |\omega_0| \leq \omega_c
\end{equation}
where $\omega_0$ and $\phi$ represent the frequency vector and the phase of illumination and $\omega_c$ is the cut-off frequency of the imaging system, and $E(r)$ is the electric field of the illumination (the intensity distribution of the illumination $I(r) = E^2(r)$); its Fourier form is expressed as:
\begin{equation}
\widetilde{E}(\omega) = \delta(\omega - \omega_0)e^{-i\phi} + \delta(\omega + \omega_0)e^{i\phi}, |\omega_0| \leq \omega_c
\end{equation}
where $\widetilde{E}(\omega)$ refers to the Fourier spectrum of the $E(r)$, $\delta(\omega)$ is a pulse function, which can be considered a point light source. The central idea of our pattern generation strategy is to place two point light sources at the conjugate Fourier plane of the illumination path in order to generate two-beam interference patterns at the sample plane.

\begin{figure}
\centering
\includegraphics[width=100mm]{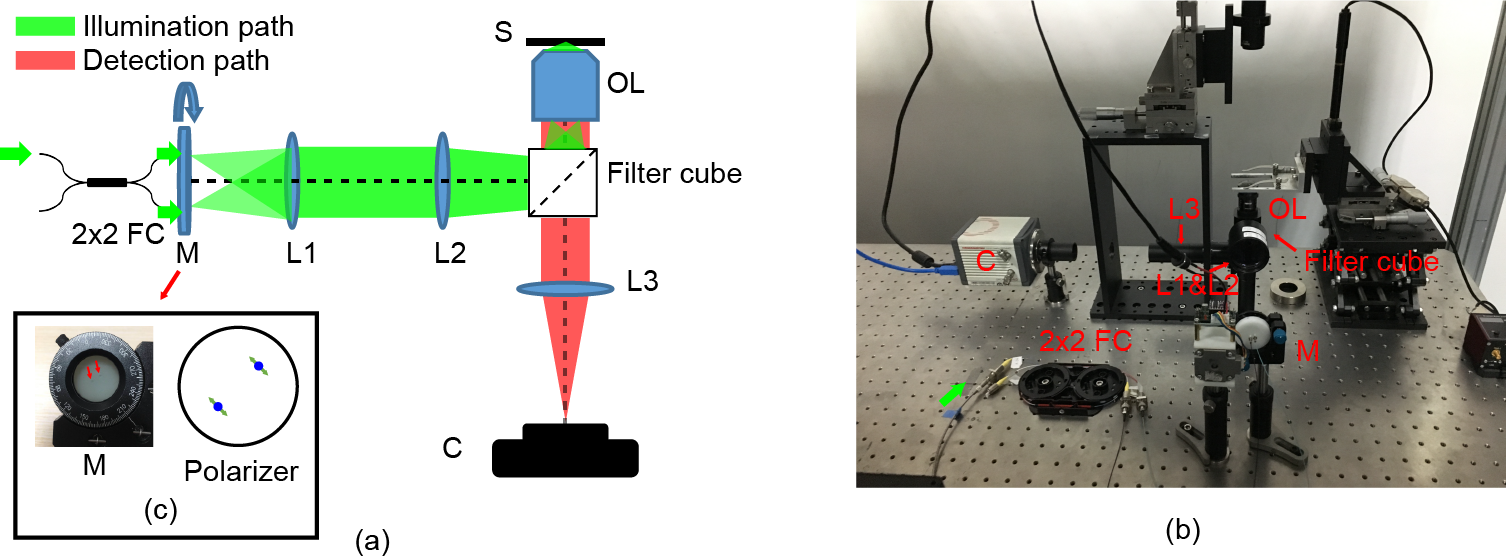}
\caption{Optical setup. (a) shows a schematic of the homemade SIM system using $2 \times 2$ fiber coupler. The output tips of the fiber coupler, $FC$, were inserted into the fiber mount with rotation function, $M$. And two lenses, $L_1$ and $L_2$, were used to collimate the light beams and project two coherent point sources at the back aperture of the objective lens, $OL$; in other words, $M$ was placed at the conjugate Fourier plane of the focal plane of the OL. The filter cube included three filter: an excitation filter, an emission filter, and a dichroic mirror. Thus two-beam-interference illumination was achieved at the sample plane, $S$. The fluorescent signal went through a tube lens, $L_3$, and was then captured by a sCMOS camera, $C$. (b) shows a picture of the optical setup in the experiment. (c) shows the structure of the fiber mount, $M$. We added a 3D printed component (the white part shown in the picture) to a nested rotation mount with a linear polarizer. The output pinholes were pointed out with red arrows. The polarization of the light passing through the polarizer was shown in the right panel, where the blue dots represented the position of those two fiber tips.}
\label{fig: setup}
\end{figure}

Figure \ref{fig: setup} (a) shows the optical setup of our homemade SIM system. Here, a $2 \times 2$ fiber coupler (TN532R5F2, THORLABS, US) with 50:50 coupling ratio was used to split an incident light source into two light beams with the same intensity. The fiber cores of output tips were $2.5 \mu m$, which could be considered two coherent point light sources. The light source we used in our setup was a fiber coupled solid-state laser (LC-651A Laser Combiner, ColdSpring, China) with $532 nm$ wavelength and $100 mW$ power.
A 3D printed fiber mount $M$, shown in Figure \ref{fig: setup} (c), was used to mount the two output fiber tips in parallel, which were then placed at the conjugate Fourier plane of the illumination path. It included three parts: a 3D printed component used to insert fiber tips in parallel, a nested rotation mount with a stepper motor and a linear polarizer that can rotate with the fiber tips. The polarization of the light beams passing through $M$ was set as shown in the right panel of Figure \ref{fig: setup} (c) in order to get better fringe illuminations.
With two lenses ($L_1$ $150 mm$, $L_2$ $100 mm$), those two point light sources were projected onto the back aperture of the objective lens (UPlanFLN, $60\times/1.25$, Olympus, Japan) while being interfered at the sample plane.
The fluorescence signal was captured by a sCMOS camera (ORCA-Flash 4.0 V2, HAMAMATSU, Japan) with a $200 mm$ tube lens, $L_3$. By rotating the fiber mount, $M$, SI images with different pattern orientations were obtained. It should be noted that the polarizer and the fiber tips were rotated at the same time to maintain the polarization of the coherent light beams.

In the proposed SIM system, the fiber coupler was a passive device which could not modulate the phase of the output light. In other words, only one SI image per orientation was captured at one time, which means the sum of all raw images could not be considered a uniform illuminated sample image. Thus, nether the most common SIM image processing algorithms, which are based on separating high-frequency components in the Fourier domain \cite{Gustafsson2000,Lal2016} or blind reconstruction algorithms \cite{Mudry2012,Dong2014a,Cao2017} were suitable in this case. As was recently reported in \cite{Ingaramo2014}, a joint RL deconvolution algorithm can combine multiple images collected under different excitation patterns; these could then be directly applied to our image recovery processing.

We then briefly refer to the SIM reconstruction based on jRL deconvolution algorithm. When using two-beam-interference patterns to illuminate the sample, the captured SI image can be expressed as follow:
\begin{equation}
  M(r) = [S(r)I(r)] \otimes h(r) = [S(r) (1 + cos(\omega_0 r + 2\phi)] \otimes h(r),
\end{equation}
where $M(r)$ is the measured fluorescent signal, $S(r)$ is the intensity distribution of the sample fluorescence, and $h(r)$ is the point spread function(PSF) of the imaging system. In conventional SIM image processing, by using convolution theorem, it is always illustrated in Fourier domain:
\begin{equation}
  M(\omega) = [S(\omega) + S(\omega-\omega_0)e^{-i2\phi} + S(\omega + \omega_0)e^{i2\phi}] \times H(\omega).
\end{equation}
High spatial frequency components are separated by a matrix inversion and at least nine raw images are required to get an isotropic resolution enhancement \cite{Lal2016}. However, according to Strohl's work \cite{Str?hl2017}, using a maximum likelihood approach, the super-resolution image recovery problem can be reformulated as an minimization problem:
\begin{equation}
  e = \min_{\tilde{e}}[\sum_{n=1}^{N}(m_n - (I_n \times \tilde{e}) \otimes h)^2],
\end{equation}
where $e$ is an estimate of the high-resolution image, $m_n$ are the measured raw images, and $I_n$ are the illumination patterns of each measured data.  A a joint Richardson-Lucy (jRL) deconvolution algorithm \cite{Ingaramo2014,Strohl2015} can solve this minimization problem. The iterative regularized reconstruction procedure could be expressed as follow:
\begin{equation}\hat{m} = h \otimes (e_i \times I) + b \end{equation}
\begin{equation}r = m/\hat{m}\end{equation}
\begin{equation}\hat{e}_{i+1} = \hat{e}_i \times [h \otimes (r \times I)]/[h \otimes(ones \times I)]
\end{equation}
where $e_i$ is an estimate of the real fluorescence distribution of the sample for iteration $i$, $h$ is the point spread function (PSF) of the imaging system, and $b$ is the random noise. Furthermore, $m$ represents the raw SI data of our measurements, $\hat{m}$ is the estimate SI data from $e_i$; $I$ represents the illumination profile of the raw SI data $m$, and $ones$ is a ones array with the same size as $m$. Here, $\times$ and $/$ represent point-wise multiplication and division and $\otimes$ represents the convolution operation. The essence of the resolution enhancement is to minimize the difference between the real raw data and artificial raw data. It should be noted that the algorithm based on jRL deconvolution can be used in all kind of SIM methods. Strohl's work shown that only three SI images can achieve similar resolution as the conventional SIM, though the mixing matrix of 3-frame SIM data without phase shift was not uniquely invertible which brings artifacts in image reconstruction. However, in our method, the mode of fiber mounts' rotation was continue and the SI patterns could be projected at multiple orientations ($>3$) which can offset those artifacts.

\section{Result}

In the experiment, we set the distance of two fiber tips to $8 mm$. Those two point sources were projected onto the back aperture of OL with a distance of $5.33 mm$. Considering the OL we used in our setup ($60 \times$, $1.25 NA$), the diameter of OL's back aperture was $7.50 mm$, which means the SI patterns were set to $71.1\%$ of the maximal spatial frequency of the $532 nm$ excitation light. And the power of laser source was set to be $100 mW$.

\begin{figure}
\centering
\includegraphics[width=100mm]{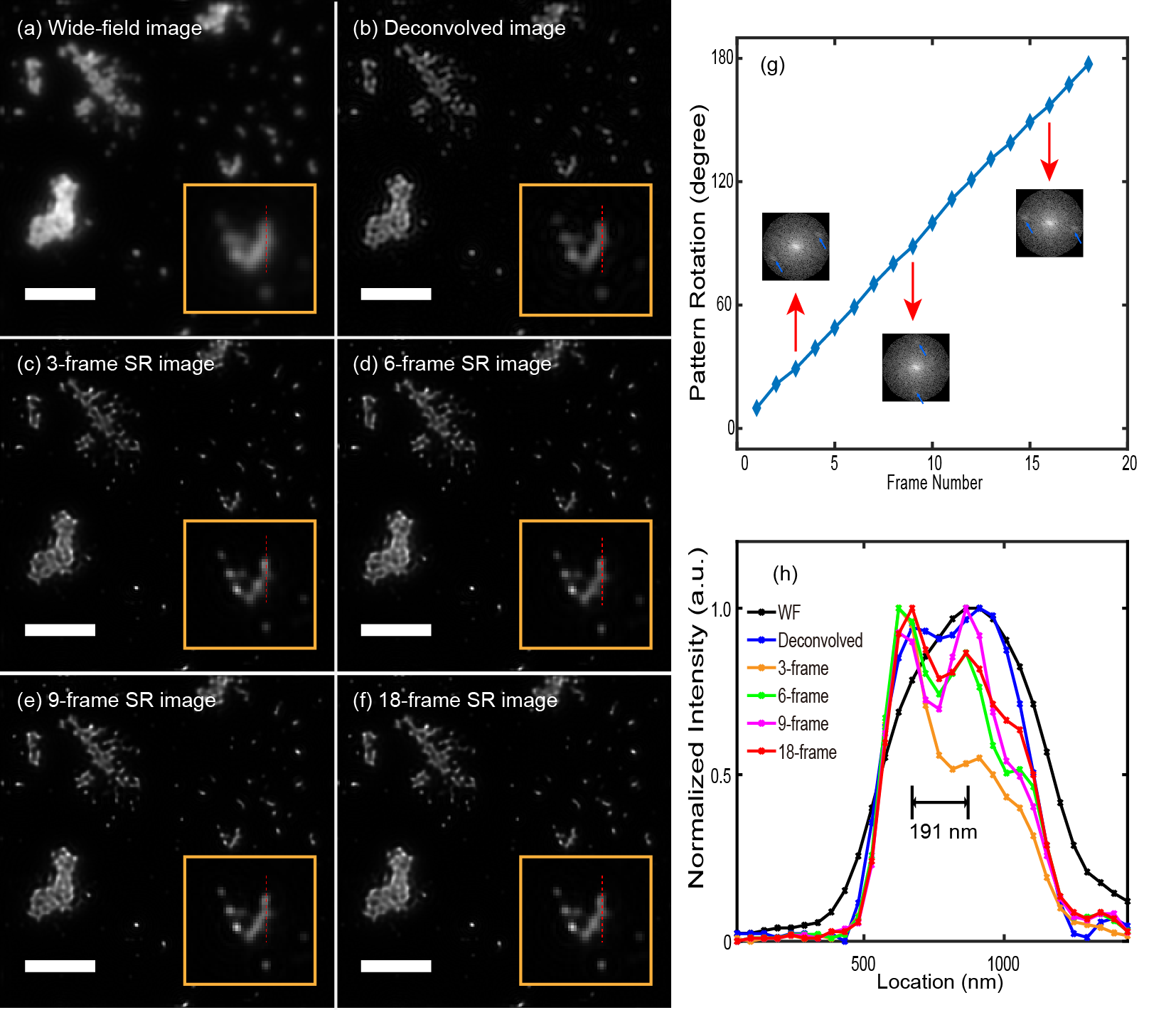}
\caption{Fluorescent microspheres imaging result in different reconstruction mode. (a) Wide-field image was captured by blocking one of the fiber tips. (b) Wide-field image deconvolved using jRL algorithm. (c)-(f) SR images are separately reconstructed from raw data with different rotation step of $60^{\circ}$ (3-frame), $30^{\circ}$ (6-frame), $20^{\circ}$ (9-frame), and $10^{\circ}$ (18-frame). (g) The rotation distribution of total eighteen raw SI images, and three spatial frequency spectrums are shown. (h) The normalized intensity distribution along the broken red line in the zoom area is shown. As shown, all the SR images distinguish two beads and achieved 191 nm resolution which is bypass the diffraction limit. The scale bars are $5 \mu m$.}
\label{fig: result1}
\end{figure}

Following this, we first imaged some fluorescent microspheres with a diameter of $100 nm$ (TetraSpeck, Thermo Fisher Scientific) on our SIM system. We rotated the fiber mount by a step of 10 degrees, and 18 raw SI images with no phase shift were captured by the sCMOS camera in 9 seconds (The exposure time of each SI image was $200 ms$). Figure \ref{fig: result1} (g) shows the pattern rotation distribution of those raw SI images ($10^{\circ} \sim 180^{\circ}$). In order to show the resolution enhancement, we separately reconstructed super-resolution images from 3-frame (rotation step: $60^{\circ}$), 6-frame (rotation step: $30^{\circ}$), 9-frame (rotation step: $20^{\circ}$), and 18 frames (rotation step: $10^{\circ}$) SI data, shown in Figure \ref{fig: result1} (c-f). As we mentioned above, in our setup, uniform-illuminated sample image could not be obtained by summing up all of the raw images; we blocked one of the fiber tips to produce a wide-field sample image, shown in Figure \ref{fig: result1} (a). And Figure \ref{fig: result1} (b) shows its deconvolved image using jRL deconvolution. All the deconvolved images shown in Figure \ref{fig: result1} were deconvolved by using fifty jRL iterations.  According to the intensity distribution shown in Figure \ref{fig: result1} (h), all the SR images distinguish two neighboring beads at a distance of 191 nm, which achieved resolution beyond the diffraction limit. It should be noticed that with more raw SI images at multiple pattern orientations, better image reconstruction quality could be achieved, and in our experience, size to nine frames are enough to achieve the conventional SIM SR effect.

\begin{figure}
\centering
\includegraphics[width=90mm]{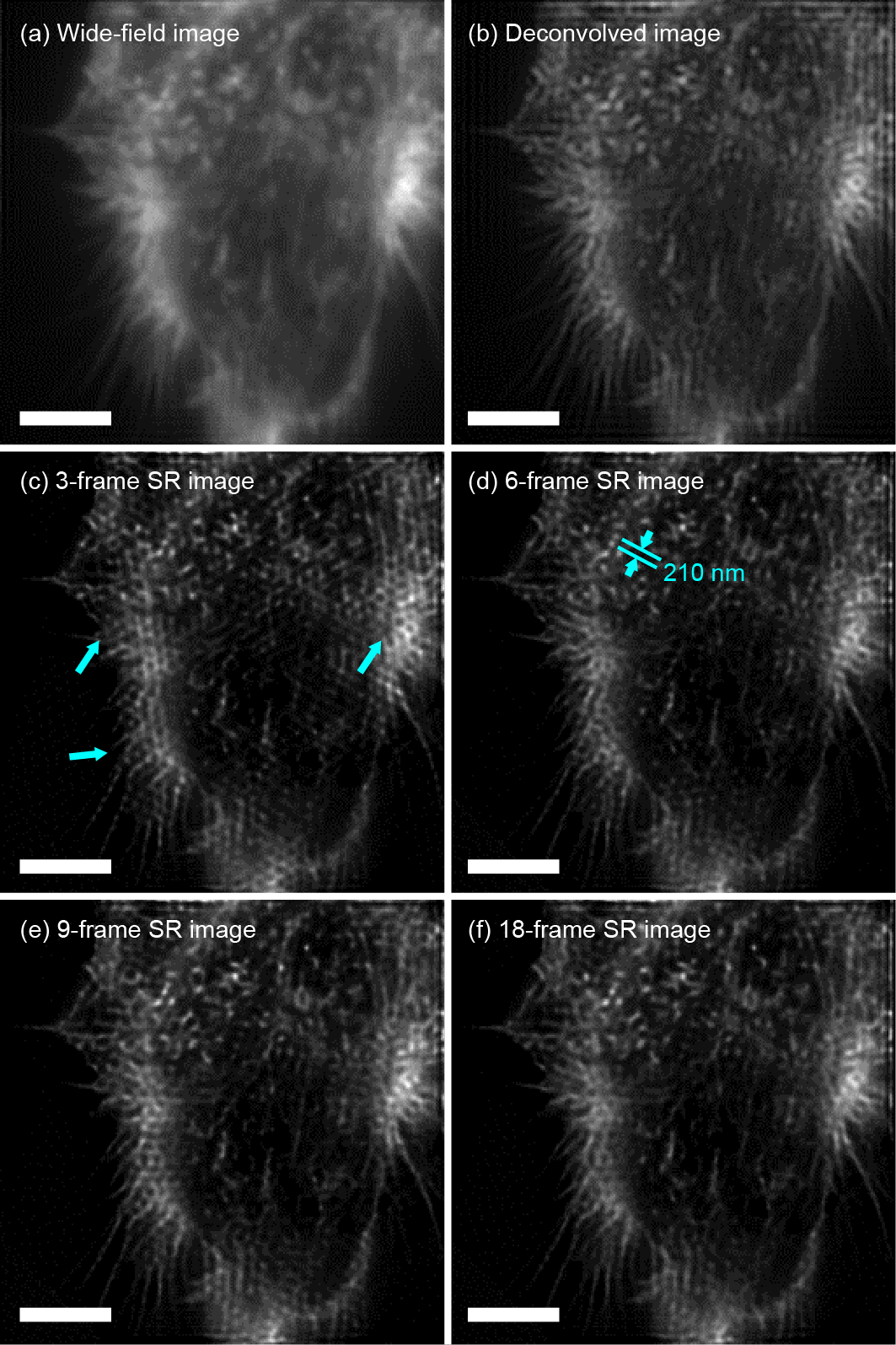}
\caption{Image of CaSki cells. (a) Wild-field image was captured by blocking one the fiber tips. (b) Wide-field image deconvolved using jRL algorithm. (c)-(f) SR images are reconstructed from 3-frame, 6-frame, 9-frame, and 18-frame data using jRL algorithm. Fifty jRL iterations were used in each deconvolved image. The scale bars are $5 \mu m$.}
\label{fig: result2}
\end{figure}

To further verify our method, we then imaged some Rhodamine Phalloidin labeled CaSki cells. The imaging result was shown in Figure \ref{fig: result2}. We used the same system setting and the exposure time of the camera was set to be $300 ms$. Eighteen raw SI images without phase shifting were captured in $10.8 s$. We also reconstructed the SR images from 3-frame, 6-frame, 9-frame, and 18-frame data using jRL algorithm, shown in Figure \ref{fig: result2} (c)-(f). As shown, better reconstruction quality could be achieved from raw data with more SI images in multiple pattern orientations, and we can see some artifacts in Figure \ref{fig: result2}, pointed by blue arrows, produced by SIM reconstruction with no phase shift as we mentioned earlier.

\section{Discussion}

\begin{figure}
\centering
\includegraphics[width=90mm]{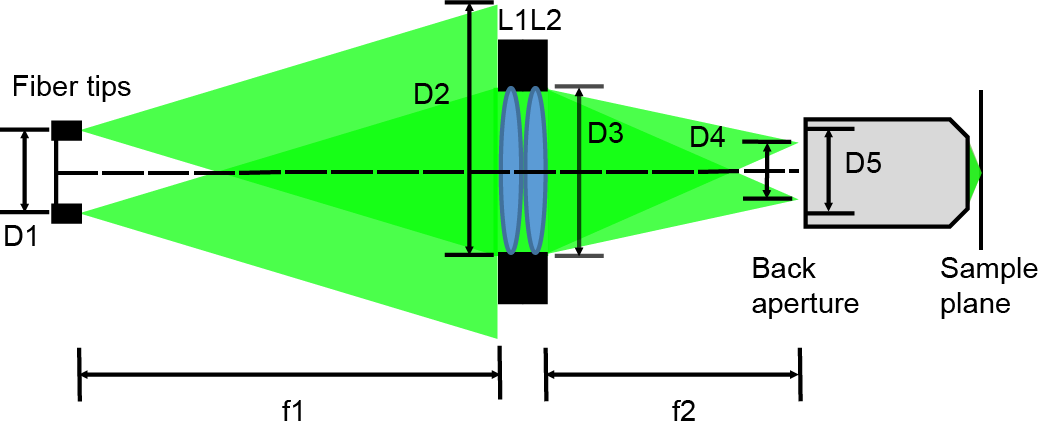}
\caption{The illumination path of the SIM setup. Two fiber tips were placed in parallel at the focal plane of $L_1$. $D_1$ is the distance between two optical fiber tips; $D_2$ is the diameter of the light field in front of $L_1$; $D_3$ is the diameter of optical lenses; $D_4$ is the distance between two point light sources at the back aperture of the objective lens whose diameter is $D_5$. $f_1$ and $f_2$ are the focal lengths of $L_1$ and $L_2$. }
\label{fig: ef}
\end{figure}

Figure \ref{fig: ef} shows the illumination path of the proposed SIM system. $D_1$ represents the distance between two fiber tips placed in parallel. $L_1$ with focal length of $f_1$ was used to collimate the light beams outputted by the fiber tips and $L_2$ with focal length of $f_2$ was used to project two point light sources onto the back aperture of the objective lens. The distance between those two point sources, $D_4$ directly corresponded to the spatial frequency of the interference pattern at the sample plane, where $D_4 = D_1 f_2 / f_1$. Hence the SR factor of the system could be expressed as follow,
\begin{equation}
  F_{SR} = D_4 / D_5 + 1 = \frac{D_1 f_2}{2 f_1 NA_{OL} f_{OL}} + 1,
\label{eq: SR factor}
\end{equation}
where $F_{SR}$ means the SR factor, $NA_{OL}$ and $f_{OL}$ represent to the numerical aperture and the focal length of the objective lens, and $D_5$ represents to the diameter of the OL's back aperture which corresponded to the maximum spatial frequency that the OL could detect. When those two point sources are projected close to the edge of the back aperture of the OL, the imaging system could achieve maximum resolution enhancement with the SR factor equals two.
According to the equation mentioned above, the SR factor (or the frequency of the illumination pattern) was determined by the distance between two fiber tips ($D_1$), the focal length of lenses ($f_1$ and $f_2$), and the optical property of the objective lens ($NA_{OL}$ and $f_{OL}$). In the experiment, when the lenses and the objective lens was set, frequency modulation of the pattern could be achieved by changing different fiber mounts with different $D_1$. Typically, in a SIM system with a certain objective lens, the spatial frequency of the SI pattern must to be close to the cut-off frequency of the imaging system to produce the best resolution enhancement, which means, in daily use, the fiber mount is not required to be changed frequently.

As we mentioned above, many methods have been proposed to create home-build SIM setups. A grating with a linear translation stage and a rotation stage was used in Wang's work \cite{Wang2011}, which used the diffracted $\pm 1st$ order beams to form interference patterns. Considering the speed and accuracy of the pattern generation, methods without moving part, such as spatial light modulator (SLM) \cite{Chang2009,Lu-Walther2015,Song2016}, digital micro-mirror device (DMD) \cite{Dan2013}, and diffractive optical elements (DOEs) \cite{Rodriguez2008}, are much preferred in building a high performance SIM system due to their high modulation speed and stability. For example, a two-beam SIM acquisition rate of 79 fps (the raw frame rate was 714 fps) was achieved using a SLM and a sCMOS camera working in continuous rolling shutter mode, proposed by Song el al. \cite{Song2016}. Compared with those methods, our method sacrificed the acquirement speed and pattern's stability caused by fibers' rotation, in order to achieve low cost arrangement. In the experiments, the orientation modulation of the SI pattern was achieved using mechanical rotation of the 3D printed fiber mount, which caused the deformation of the fibers. This led to an uncertain phase shift and polarization change of interference pattern, which could cause the appearance of artifacts in image reconstruction. Though the mechanical rotation could offer a continuous change of pattern's orientation which can offset the artifacts in jRL deconvolution with no-phase-shift raw data, the acquisition speed of SI images was very slow (In the experiments, it takes about $10s$ to capture 18 raw images.). These issues can be mitigated by using more fiber tips pairs as a programmable coherent point-source array, thereby avoiding mechanical movement. Moreover, like SIM system based on gratings, it is not convenient to modulate the spatial frequency of the SI patterns.

However, there are two advantages of our method: high power efficiency and compact illumination path. For most of the methods we mentioned above, a grating-like structure was used to diffract the incident laser light into $\pm 1st$ order beams, which interfered at the sample plane to form SI pattern. A major drawback of them is the low light transmission caused by diffraction. In our method, as shown in Figure \ref{fig: ef}, the fiber tips were placed off-axis, while the light beams collimated by $L_1$ had a certain angle in relation to the optical axis. If we set $L_1$ and $L_2$ to be a 4F system, most light cannot enter the pupil of $L_2$ owing to the long distance between two lenses. To improve the light transmission, we set $L_1$ and $L_2$ to be as close as possible so that most of the energy could be used. Hence, the power efficiency was approximately determined by the light entering the pupil of the lenses, which could be expressed as
\begin{equation}
P = \left\{
  \begin{array}{lr}
    100 \%, & D_3 \geq D_1 + D_2,\\
    (\frac{D_3+D_2-D_1}{2 D_2})^2, & D_2 - D_1 \leq D_3 < D_1 + D_2,\\
    (\frac{D_3}{D_2})^2, & D_3 < D_2 - D_1,
  \end{array}
\right.
\label{eq: power efficiency}
\end{equation}
where $D_2 = 2 tan(\alpha) f_1$, $\alpha = arcsin(NA_{fiber})$, and $P$ represents the power efficiency. Here we assume the light outputted by the fiber tips is uniform and the transmitted energy is proportional to the transmitting area (the pupil of the lenses). We noted that we only calculated the power efficiency from the fiber tips to the back aperture of the objective lens; we ignored the insertion loss of the optical fiber, which was about $3 dB$. In the experiment, the diameter of the lenses was $50 mm$, $D_3$, and the $NA$ of the fiber tips was $0.13$. When the focal length of the $L_1$ was set to $150 mm$, approximately $100 \%$ of the light energy crossed the illumination path and projected onto the sample plane. Thus, high power efficiency could be achieved in our method.
Moreover, unlike the grating-structure-based setups which need a 4F system to let only $\pm 1$ order beams to pass, or the DOEs-based setup which needs one DOE per pattern orientation, our method directly placed two point light sources (two fiber tips) at the conjugate Fourier plane of the sample plane, which allows shortening illumination path, as shown in Figure \ref{fig: setup} (b). Thus, our method could be easily applied to a commercial microscope due to its low cost and simplicity.

\section{Conclusion}

In summary, we have proposed a compact and low-cost SIM system using a $2 \times 2$ fiber coupler. Two output fiber tips, mounted on a rotatable 3D-printed fiber mount, were placed in parallel at the conjugate Fourier plane of the illumination path, which generated two-beam interference excitation patterns at the sample plane. Resolution-enhanced images of fluorescent beads and CaSki cells were recovered from the no-phase-shift raw SI images with a joint Richardson-Lucy deconvolution algorithm. Our proposed method offers a cheap and simple way to build a SIM setup with high power efficiency.

In future work, more coherent point light sources (like more fiber couplers) could be added as an array placed at the conjugate Fourier plane, thereby replacing mechanical movement with a programmable operation of the point light source array. Thus, the reconstruction artifacts in the SR images could be reduced and the acquisition speed of the system could be improved. It should be noted that, SIM image processing based on joint Richardson-Lucy deconvolution with no-phase-shift data would produce artifacts when the dataset was small (like 3 SI raw images). The reconstruction images could be improved by adding more SI images in different orientations or adding one uniform-illuminated (wide-field) image. In optical setup, a uniform illumination could simply achieved by adding a single channel point light source at the array like light-emitting diodes (LEDs) or laser diodes (LDs). Furthermore, several 4-frame reconstruction algorithms \cite{Orieux2012,Dong2015,Lal2016a,Meiniel2017}, based on one wide-field image and three SI images, could be applied to recover SR images in future work.

\section*{Acknowledgments}
This work was supported in part by National Natural Science Foundation of China under [Grant numbers 61334008, 21804126], the Instrument Developing Project of the Chinese Academy of Sciences [Grant number YZ201402], and the National Key Research and Development Program of China [Grant number 2018YFF0214900].

\section*{References}

\bibliography{refbibfile}

\end{document}